\documentclass[prl,twocolumn,showpacs,superscriptaddress]{revtex4-1}

\usepackage{graphicx,bm,amssymb,amsmath}
\usepackage{color}
\usepackage{amsmath}

\begin{document}

\title{Active Jamming: Self-propelled soft particles at high density}

\author{Silke Henkes}
\affiliation{Physics Department, Syracuse University, Syracuse, NY 13244, USA}
\author{Yaouen Fily}
\affiliation{Physics Department, Syracuse University, Syracuse, NY 13244, USA}
\author{M. Cristina Marchetti}
\affiliation{Physics Department, Syracuse University, Syracuse, NY 13244, USA}
\affiliation{Syracuse Biomaterials Institute, Syracuse University, Syracuse, NY 13244, USA}

\date{\today}

\begin{abstract}
We study numerically the phases and dynamics of a dense collection of self-propelled particles with soft repulsive interactions  in two dimensions. The model  is motivated by 
recent in vitro experiments on confluent monolayers of 
migratory epithelial and endothelial cells. The phase diagram exhibits a liquid phase with giant number fluctuations at low packing fraction $\phi$ and high self-propulsion speed $v_0$ and a jammed phase at high $\phi$ and low $v_0$. The dynamics of the jammed phase is controlled by the low frequency modes of the jammed packing.
\end{abstract}

\maketitle

How do collections of active particles behave in very dense situations? What are the mechanical properties of the ensuing materials? The answers to these questions are fundamentally important for a wide range of physical and biological systems, from tissue formation~\cite{Angelini2011,Angelini2010,Trepat2009,Poujade2007,Petitjean2010} and vibrated granular materials~\cite{Narayan2007,Deseigne2010} to the behavior of packed crowds~\cite{Helbing2007}.

The name ``active matter''  refers to soft materials composed of many interacting units that individually consume energy and collectively generate motion or mechanical stress. Examples range from bacterial suspensions to epithelial cell layers and flocks of birds. The  phases of active matter have been studied extensively since the seminal work of Vicsek et al~\cite{Vicsek1995}. Self-propelled  particles have a  polarity provided by the direction of self-propulsion. In the presence of noisy polar aligning interactions,  they order into a moving state at high density or low noise~\cite{Gregoire2004,Chate2008}. The ordered  state has giant number fluctuations~\cite{Chate2006,Narayan2007,Deseigne2010} and a rich spatio-temporal dynamics. Continuum theories have been formulated for these systems and provide a powerful tool for understanding the generic aspects of their behavior~\cite{TonerTuRamaswamy2005}. While the low density phase of  various  models of self-propelled particles is comparatively well understood, much less is known about the high density phase.

In a separate development, much effort has been devoted to the study of passive thermal and athermal granular matter. These systems undergo a transition between a flowing, liquid-like state at low density or high temperature and a glassy state~\cite{Liu1998,Berthier2009}. Near the glass transition, the relaxation is controlled by dynamical heterogeneities, consisting of spatially and temporally correlated collective rearrangements of particles~\cite{Weeks2000}. In the zero-temperature limit, soft repulsive disks undergo a jamming transition to mechanically stable state at $\phi=0.842$ in two dimensions~\cite{O'Hern2003}. The elastic properties of the jammed state are determined by an excess number of low frequency modes~\cite{Silbert2005} which are also closely linked to the large-scale rearrangements that microscopic packings undergo when strained~\cite{Maloney2006} or  thermalized~\cite{Brito2007}.

Recent in vitro experiments on confluent monolayers of migratory epithelial and endothelial cells have revealed displacement fields and stress distributions that strongly resemble both dynamical heterogeneities of glasses and the soft modes of jammed packings~\cite{Angelini2011,Angelini2010,Trepat2009,Poujade2007,Petitjean2010}, and an analogy between the dynamics of these living systems and that of  glassy materials has been proposed~\cite{Angelini2011}. 
Migratory cells have been successfully modeled with soft polar particles at low densities~\cite{Szabo2006} as well as through a mechanical agent model at high densities~\cite{Bindschadler2007}. 

Motivated by the experiments on dense migrating cell layers, in this paper we consider a model that combines the polar alignment mechanism of self-propelled particles  with the repulsive soft disk interactions that control  jamming and characterize the  phases and their dynamical properties. The model is based on that of Vicsek~\cite{Vicsek1995} and was introduced some time ago by Szabo et al.~\cite{Szabo2006}   to describe the collective dynamics of crawling cells. These authors focused  on the swarming behavior of collections of cells below close packing,  with the goal of understanding the collective dynamics as a  flocking transition. Here we explore the full phase diagram of a confined system. We show the existence of an active jammed state at high density with glassy dynamics and fluctuations of the type seen in confluent cell layers.

\begin{figure}
\centering
\includegraphics[width=0.49\columnwidth,trim = 30mm 30mm 28mm 28mm,clip]{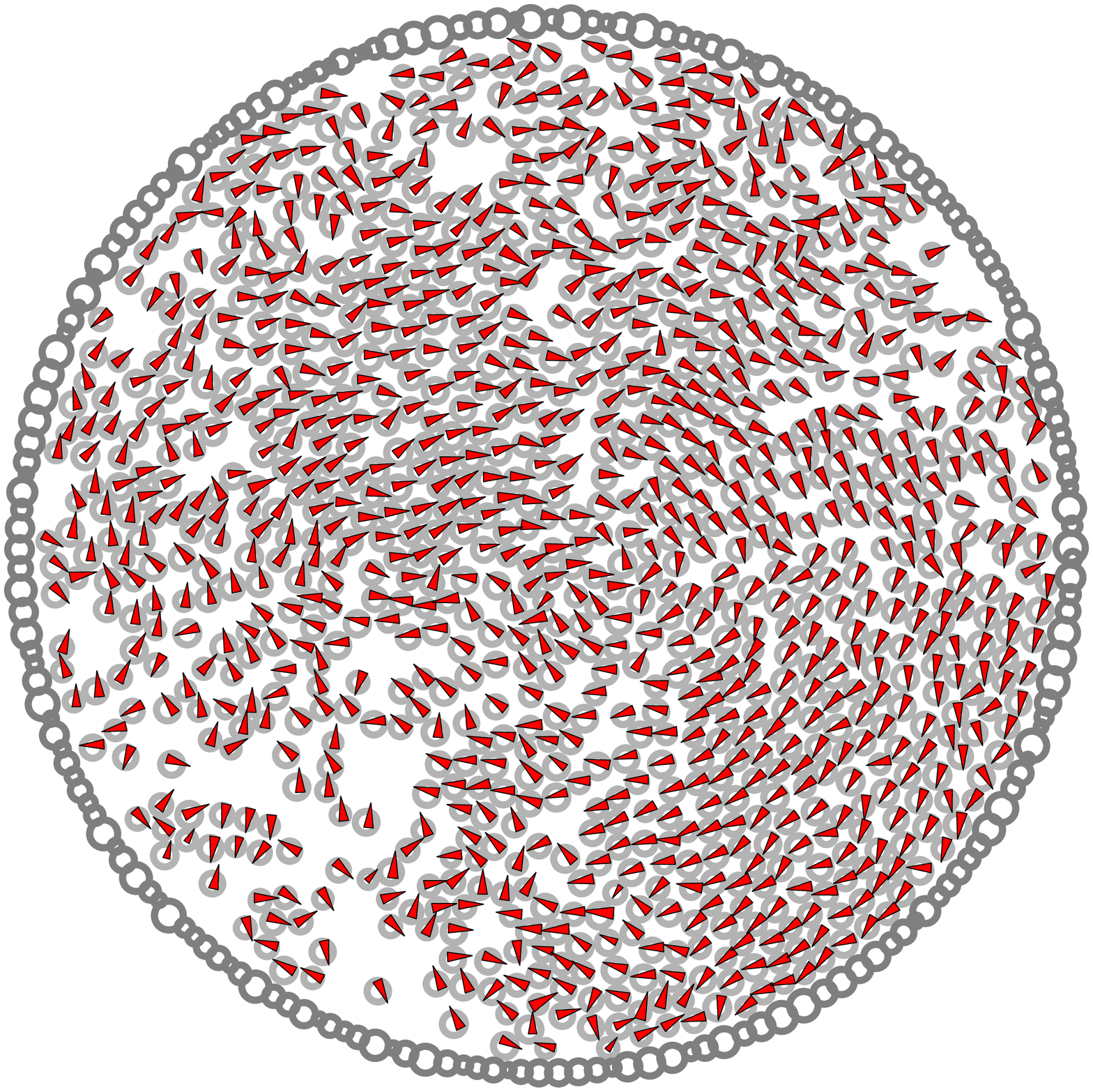}
\includegraphics[width=0.49\columnwidth,trim = 30mm 30mm 28mm 28mm,clip]{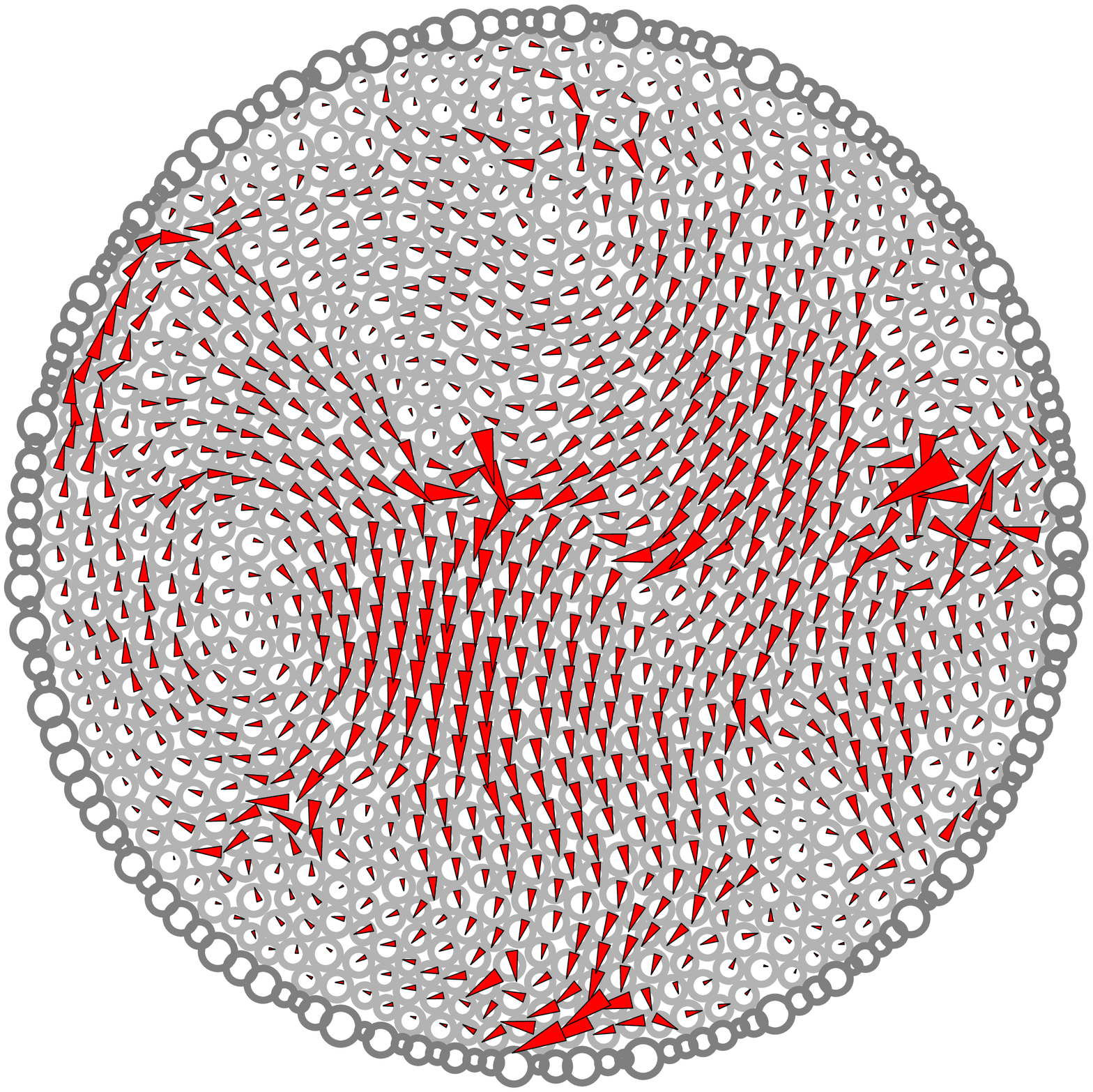}
\caption{(color online) Sample snapshots of the system in the liquid phase ($\phi=0.6$, left) and in the jammed phase ($\phi=0.95$, right) for $v_0=0.025$. The glued boundary is shown in dark grey. The arrows (red online) represent the instantaneous velocity field, with $v=v_{0}$ corresponding to an arrow of length 1 in units of the particle diameter. Please see our supplementary materials for movies of these two runs.}
\label{fig:sysplot}
\end{figure}

We consider  $N_t$ polar disks of radii $a_i$ and denote by  ${\bf r}_i$ the instantaneous positions of the disks' centers.
The disks' polarity is defined by unit vectors $\hat{\bf n}_i=\cos\psi_i\ \hat{\bf x}+\sin\psi_i\ \hat{\bf y}$ that fix a direction on each disk.
The dynamics is described by coupled equations for the translational and angular degrees of freedom, given by ~\cite{Szabo2006} 
\begin{align}
& \dot{\mathbf{r}}_{i} = v_0 \hat{\mathbf{n}}_{i} + \mu \sum_{j=1}^{z_{i}} \mathbf{F}_{ij} \;, \nonumber \\
& \dot{\psi}_{i} = \frac{1}{\tau} \left( \theta_i-\psi_i \right) + \eta_{i}\;, \quad \left \langle \eta_{i} \eta_{j} \right \rangle = \sigma^{2} \delta_{ij} \delta(t-t') \;.
\label{eq:model}
\end{align}
The dynamics of ${\bf r}_i$ is overdamped with  two sources of motion: a self propulsion velocity of constant magnitude $v_0$ directed along $\hat{\bf n}_i$, and the velocity arising from repulsive contact forces ${\bf F}_{ij}$ on particle $i$ from its  $z_i$ neighbors, with  $\mu$ a mobility.
The  force between particles $i$ and $j$ is 
${\bf F}_{ij}=-k (a_i+a_j-r_{ij}) \hat{\bf r}_{ij}$ if $r_{ij}<a_i+a_j$ and ${\bf F}_{ij}={\bf 0}$ otherwise, corresponding to soft spheres.
The radii are random and uniformly distributed in $a=[0.8,1.2]$.
The dynamics of the orientation $\psi_i$ is also overdamped. It is controlled by a
torque proportional to the angle  between $\hat{\bf n}_i$ and the direction of the velocity ${\bf v}_i=\dot{\bf r}_i=v_i\ (\cos\theta_i\ \hat{\bf x}+\sin\theta_i\ \hat{\bf y})$.
In other words, the cell's polar axis  aligns with the actual velocity, with a lag time $\tau$ and
 a gaussian random noise $\eta_i$ of zero mean and variance $\sigma^2$.
 The lag time $\tau$ describes a positive feedback between cell polarity and motility. In living cells this coupling results into directed migration of the cell that can be reinforced by neighbors in a dense cell culture, playing an important role in collective cell dynamics~\cite{Szabo2010}.
 Such polarization is not, however, permanent, but rather it is actively regulated by both biochemical processes inside the cell and feedback from neighboring cells. 
 Finally, we define dimensionless quantities by scaling all lengths with the average radius $\overline{a}$ of the spheres and and all times with the lag time $\tau$. Additionally, we fix $\mu k=10$ and $\sigma=10^{-1}$.

Using this model, we perform molecular dynamics simulations with $N_t=64$ to $10000$ particles. Unless otherwise specified, we show results for $N_t=1000$ particles. To eliminate the global translational mode obtained at high density in an open system, we confine the particles to a circular box of radius $R$ with soft repulsive boundary conditions. These are implemented by ``gluing" a row of soft spheres to the box's boundary, as shown in Fig.~\ref{fig:sysplot}.
We explore the phase diagram by varying the self propulsion speed $v_0$ and the packing fraction $\phi=\sum_i a_i^2 / R^2$.

\begin{figure}
\centering
\includegraphics[width=0.49\columnwidth,trim = 5mm 0mm 25mm 15mm,clip]{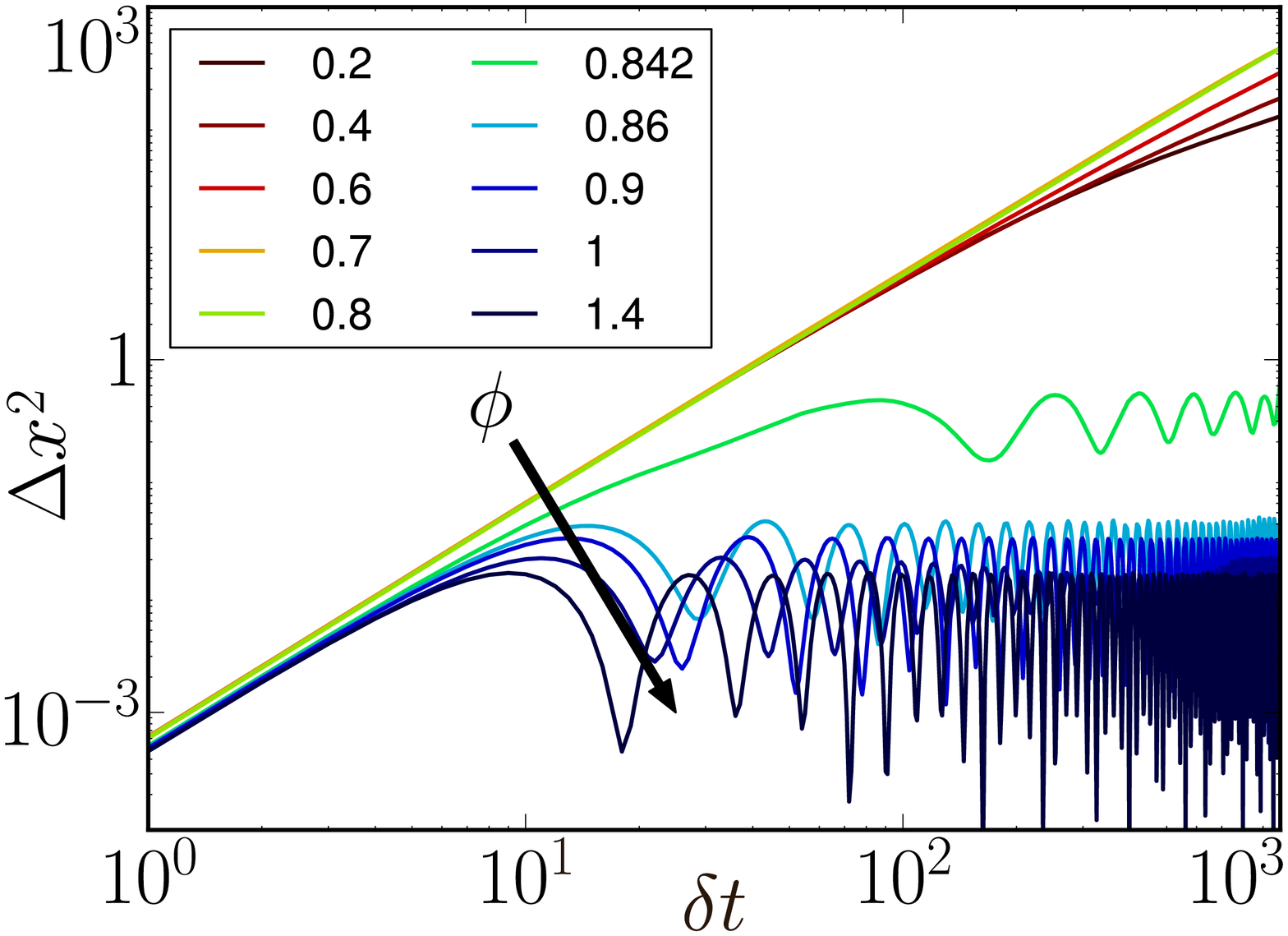}
\includegraphics[width=0.49\columnwidth,trim = 6mm 1mm 26mm 14mm,clip]{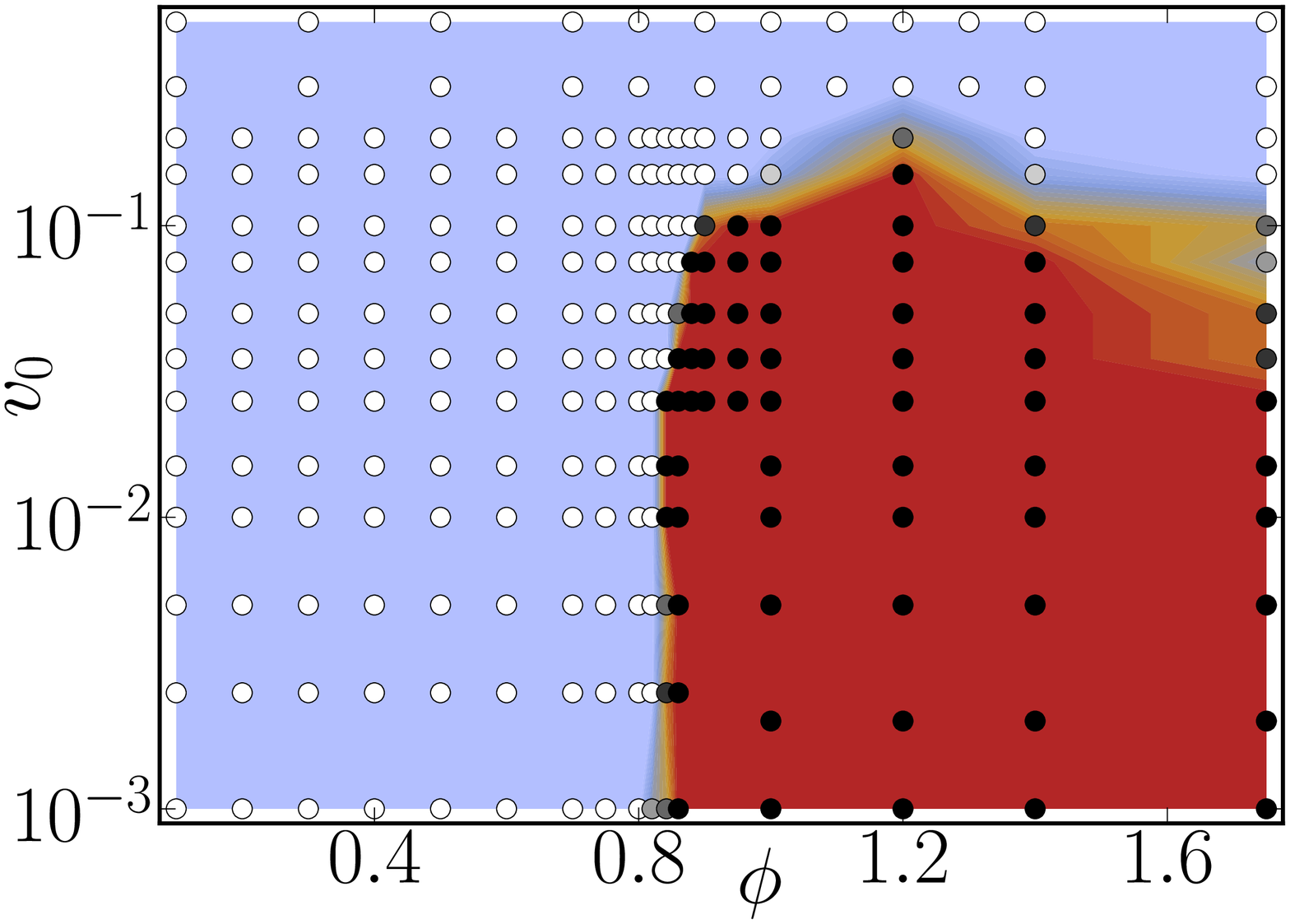}
\caption{(color online) Left: Mean square displacement vs. time as a function of density at $v_0=0.025$, showing a transition from rotational diffusion at low $\phi$, to polar alignment for $\phi<0.8$ and to the jammed state around $\phi=0.842$. Right: Phase diagram in the $\phi$-$v_0$ plane, showing the transition from the liquid state (blue online) to the solid state (red online). The dots are simulated ($\phi$,$v_{0}$)-pair shaded white to black proportional to the fraction of jammed configuration.}
\label{fig:phasediag}
\end{figure}

We first characterize the state of the system by studying the mean square displacement (MSD) of individual particles as a function of time, shown on the left side of Fig.~\ref{fig:phasediag}.
At low packing fraction or high velocity,  the MSD grows monotonically well beyond $\overline{a}$, corresponding to a flowing system.
Conversely, at high $\phi$ or low $v_0$ the MSD is bounded and smaller than $\overline{a}$, \emph{i.e.} the particles are trapped in the cage formed by their neighbors.
Typical snapshots of the system in each phase are shown in Fig.~\ref{fig:sysplot}.
At $v_0=0$, the angular degree of freedom $\psi_i$ becomes irrelevant and the problem is equivalent to the athermal jamming of soft spheres.
The transition between a flowing phase and a trapped one at very low $v_0$ is consistent with this limit; in particular, the critical packing fraction coincides with the expected value $\phi_c\approx0.842$.
By extension, we call the two active phases ``liquid" and ``jammed" respectively.
The shape of the $(\phi,v_0)$ phase diagram as inferred from the mean square displacement is shown on the right side of Fig.~\ref{fig:phasediag}.


\begin{figure}
\centering
\includegraphics[width=0.68\columnwidth,trim = 0mm 0mm 15mm 5mm,clip]{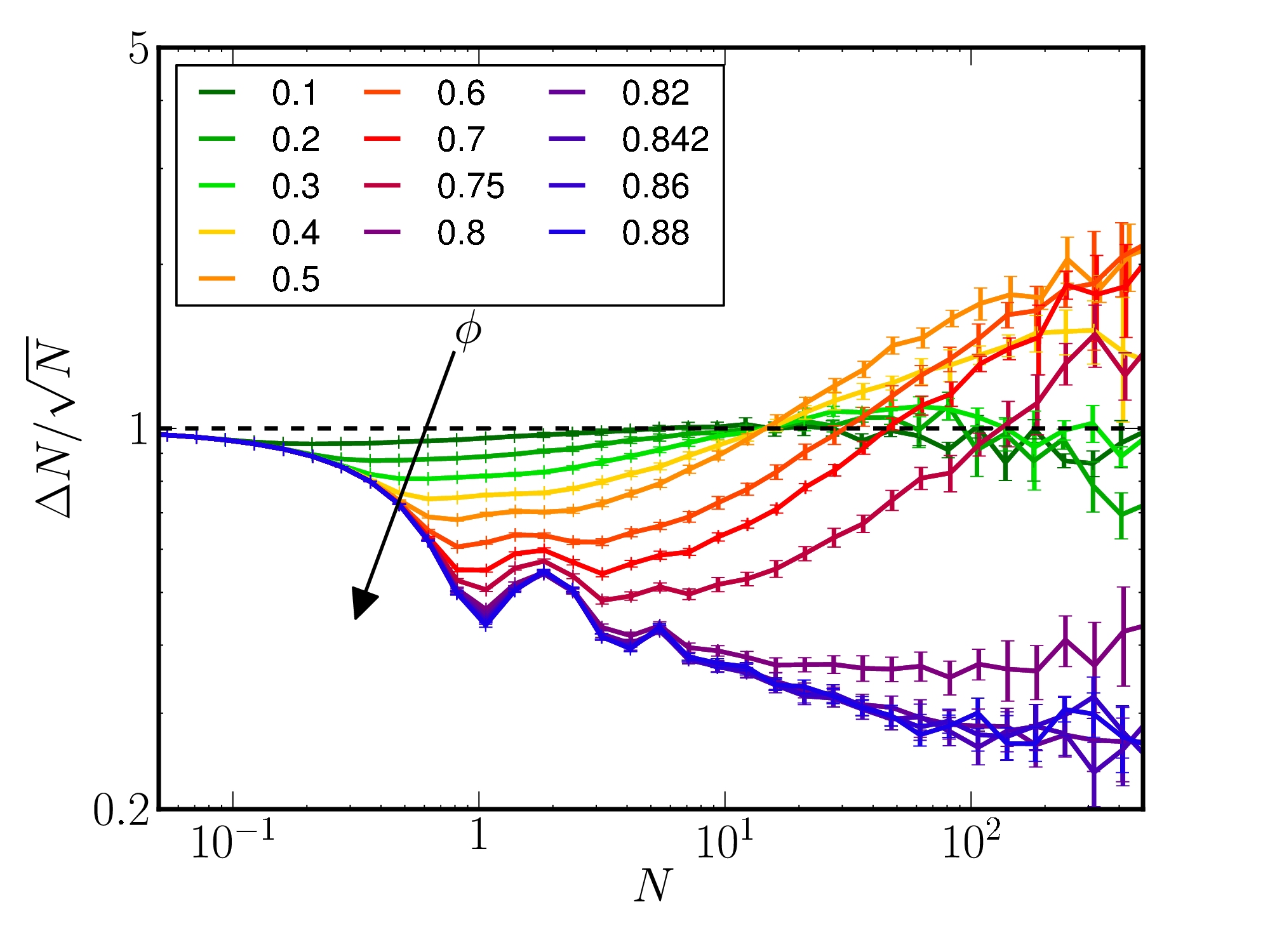}
\caption{(color online) Scaled number fluctuations for $N_t=10000$ and a cut through the phase diagram at $v_0=0.025$. We observe three regimes: gas-like fluctuations at low density (green online), giant number fluctuations at intermediate density (red online), and strongly suppressed fluctuations in the jammed phase (blue online).
The dashed line corresponds to $\Delta N/N^{1/2}=1$.
}
\label{fig:fluct}
\end{figure}

The liquid phase can be further divided by noting that the behavior of the MSD is not uniform.
At very low density, interactions are negligible and each particle independently performs a persistent random walk, with 
$\langle[{\bf r}(t)-{\bf r}(0)]^2\rangle=(4v_0^2/\sigma^2)\left[t+(2/\sigma^2)\left(e^{-\sigma^2t/2}-1\right)\right]$ and  a crossover from ballistic behavior 
$\langle[{\bf r}(t)-{\bf r}(0)]^2\rangle \sim v_0^2 t^2$  for $t<<\sigma^{-2}$ to diffusive behavior  $\langle[{\bf r}(t)-{\bf r}(0)]^2\rangle \sim (4v_0^2/\sigma^2)t$ for $t>>\sigma^{-2}$. Here $\sigma^{-2}=10^2$ and ballistic behavior is observed at all but the longest times (but shorter than the limit imposed by the box size, not shown on Fig.~\ref{fig:phasediag}), as expected for individual self-propelled particles~\cite{Howse2010}.
At intermediate density, however, clusters of aligned particles start to form and  the MSD remains ballistic at all observed times. This behavior is reminiscent of those observed in other active systems~\cite{Gregoire2004,Chate2008}.
Another signature of the symmetry breaking introduced by the active velocity in the liquid phase is the existence of  ``giant number fluctuations" ~\cite{Chate2006,Zhang2010,Narayan2007,Deseigne2010}.
The scaling of the standard deviation $\Delta N$ of the number of particles with the average number of particles $N$ in subsystems of various sizes  is shown in Fig.~\ref{fig:fluct}.
We see a transition from $\Delta N\sim N^{1/2}$, as expected in an ideal gas or in a passive termal liquid, to $\Delta N\sim N^\alpha$ with $\alpha>1/2$ at packing fraction $\phi\sim0.5$, consistent with the change of behavior observed in the MSD and with previous observations on self-propelled systems~\cite{Chate2006,Zhang2010,Narayan2007,Deseigne2010}.

In the jammed phase, we observe regular oscillations of the particle displacements around their mean positions, resulting in the oscillatory behavior of the MSD. This behavior is easily understood as that of a single particle in a harmonic well.  If we replace the force from the neighbors by the force $-k{\bf r}_i$ due to a harmonic well, the MSD can be calculated exactly. At times $t \gg \tau$, the particle performs a circular orbit of angular frequency $\omega\approx\sqrt{\mu k/\tau}$, independent of $v_0$,  and radius $R_0\approx v_0\sqrt{\tau/\mu k}$. In the jammed systems each particle is trapped in a cage determined by its neighbors and the MSD oscillates at a frequency that is independent of $v_0$, as for a single trapped particle.    The patterns observed in this phase shown in the right frame of Fig.~\ref{fig:sysplot} strikingly resemble the low-frequency modes of jammed packings and the observed oscillation frequencies are consistent with this.

For low $v_0$, within the range of linear response, we can relate the particle displacements to the soft modes of the system.  We denote by $\lbrace \mathbf{r}_{i}^{0} \rbrace$ the mean positions of all particles in a jammed packing obtained for $v_0=0$ and let $\omega_{\nu}^{2}$ and $\boldsymbol{\xi}^{\nu}$ be the eigenvalues and eigenvectors of the corresponding dynamical matrix. The assumption that  the mean positions are force-balanced and correspond to a stable jammed packing requires $\omega^{2}_{\nu}>0$ for all $\nu$. We then expand both the particle displacements $\delta \mathbf{r}_{i} = \mathbf{r}_{i}-\mathbf{r}_{i}^{0}$ and the polarization vector $\hat{\mathbf{n}}_{i} $ in the modes eigenbasis,
\begin{equation}
\delta \mathbf{r}_{i} = \sum_{\nu} a_{\nu}(t) \boldsymbol{\xi}^{\nu}_{i}\;, \hspace{0.3in}\hat{\mathbf{n}}_{i}  =  \sum_{\nu} b_{\nu}(t) \boldsymbol{\xi}^{\nu}_{i}\;. \label{eq:mode_coefficients}
\end{equation}
Our objective is to obtain linearized equations for  $a_{\nu}(t)$ and $b_{\nu}(t)$. The nonlinearities are in the equation for the cell orientation that couples the polarization angle to the direction of the velocity. This coupling through the angle, not through the vector, is a common feature of all Vicsek-type models~\cite{Chate2008}. On times long compared to the alignment time scale, $\tau$, we assume $\tau\dot{\psi}_i\ll 1$ and linearize the angular interactions by letting
\begin{equation} 
\frac{d}{dt}\left(e^{i\psi_{i}}\right)\simeq \frac{1}{\tau} \left[ e^{i\left(\theta_{v}^{i}-\psi_{i}\right)}e^{i\eta_{i} \tau}  - 1\right]e^{i \psi_{i}} \;. 
\end{equation}
In addition, we average over the angular noise using $\langle e^{i\eta_{i} \tau} \rangle = e^{-\frac{1}{4} \sigma^{2} \tau^{2}}\equiv\Delta$, and introduce a mean-field approximation for the magnitude of  local velocity by letting $v_{i}\simeq [\frac{1}{N}\sum_{i} v_{i}^{2}]^{1/2}\equiv v_{rms}$.  We then obtain linearized equations for the mode amplitudes, given by
\begin{align}
&\dot{a}_{\nu}(t) = v_{0} b_{\nu}(t)-\mu \omega_{\nu}^{2} a_{\nu}(t)\;, \nonumber \\
&\tau \dot{b}_{\nu}(t) = \frac{\Delta}{v_{rms}} \dot{a}_{\nu}(t) - b_{\nu}(t)\;, \nonumber \\
& v_{rms}^2 =\frac{1}{N}\sum_{\nu} [\dot{a}_{\nu}(t)]^{2}\;.
\label{eq:linmodes}
\end{align}
Nonlinearities now enter only through $v_{rms}$, the spatially averaged mean square particle velocity, assumed  constant below (consistently with numerics in the steady state).

The first two of Eqs.~\eqref{eq:linmodes} can be rewritten as a single second order differential equation for  $a_{\nu}$, given by
\begin{equation}
\ddot{a}_{\nu} + \frac{1}{\tau} \left[ 1- \frac{v_{0} \Delta}{v_{rms}} + \mu \tau \omega_{\nu}^{2} \right] \dot{a}_{\nu} + \frac{\mu}{\tau} \omega_{\nu}^{2} a_{\nu} =0. \label{eq:mode_oscillator} 
\end{equation}
Since $\frac{\mu}{\tau} \omega_{\nu}^{2}\geq 0$, this equation describes a damped harmonic oscillator provided the friction $f=\frac{1}{\tau} \left[ 1- \frac{v_{0} \Delta}{v_{rms}} + \mu \tau \omega_{\nu}^{2} \right]$ is non-negative. The modes amplitudes are then
\begin{align}
& a_{\nu}(t) = a_{\nu}(0) e^{- t/\tau_\nu} \cos( \omega_{\nu}' t + \theta^{0}_{\nu})\;, \label{eq:omega_theory}\\
& \omega_{\nu}' = \left[ \frac{\mu}{\tau} \omega_{\nu}^{2}-\frac{\mu^{2}}{4} \left(\omega_{\nu}^{2}-\omega_{\text{min}}^{2} \right)^{2} \right]^{\frac{1}{2}}\;, \nonumber
\end{align}
with $\tau^{-1}_{\nu} =\mu \left(\omega_{\nu}^{2}-\omega_{\text{min}}^{2} \right)/2$ and $\omega_{\text{min}}$ the smallest eigenfrequency.  For the system to reach steady-state, there need to exist finite amplitude undamped modes. Then the \emph{lowest frequency modes} of the system will be undamped, corresponding to $f=0$, which fixes  $v_{rms}$ as
\begin{equation} \mu \tau \omega_{\text{min}}^{2} = 1 - \frac{v_{0} \Delta}{v_{rms}}. \end{equation}
An important result is that
at long times the steady-state dynamics  is dominated by undamped oscillations corresponding to the lowest-frequency modes associated with the jammed packing defined by the mean particle positions.
\begin{figure}
\centering
\includegraphics[width=0.50\columnwidth,trim = 1mm 0mm 15mm 5mm,clip]{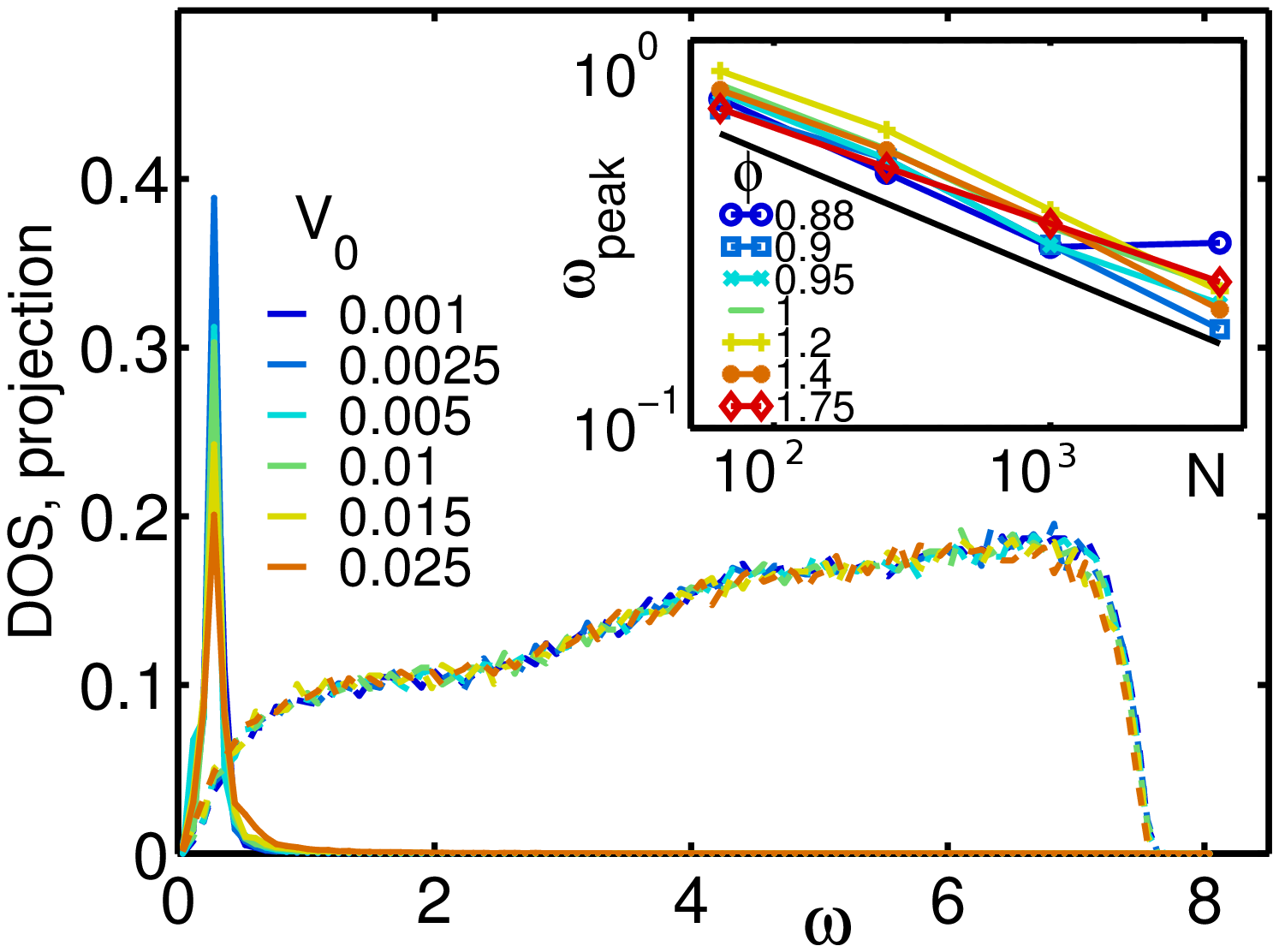}
\includegraphics[width=0.47\columnwidth,trim = 8mm 0mm 15mm 5mm,clip]{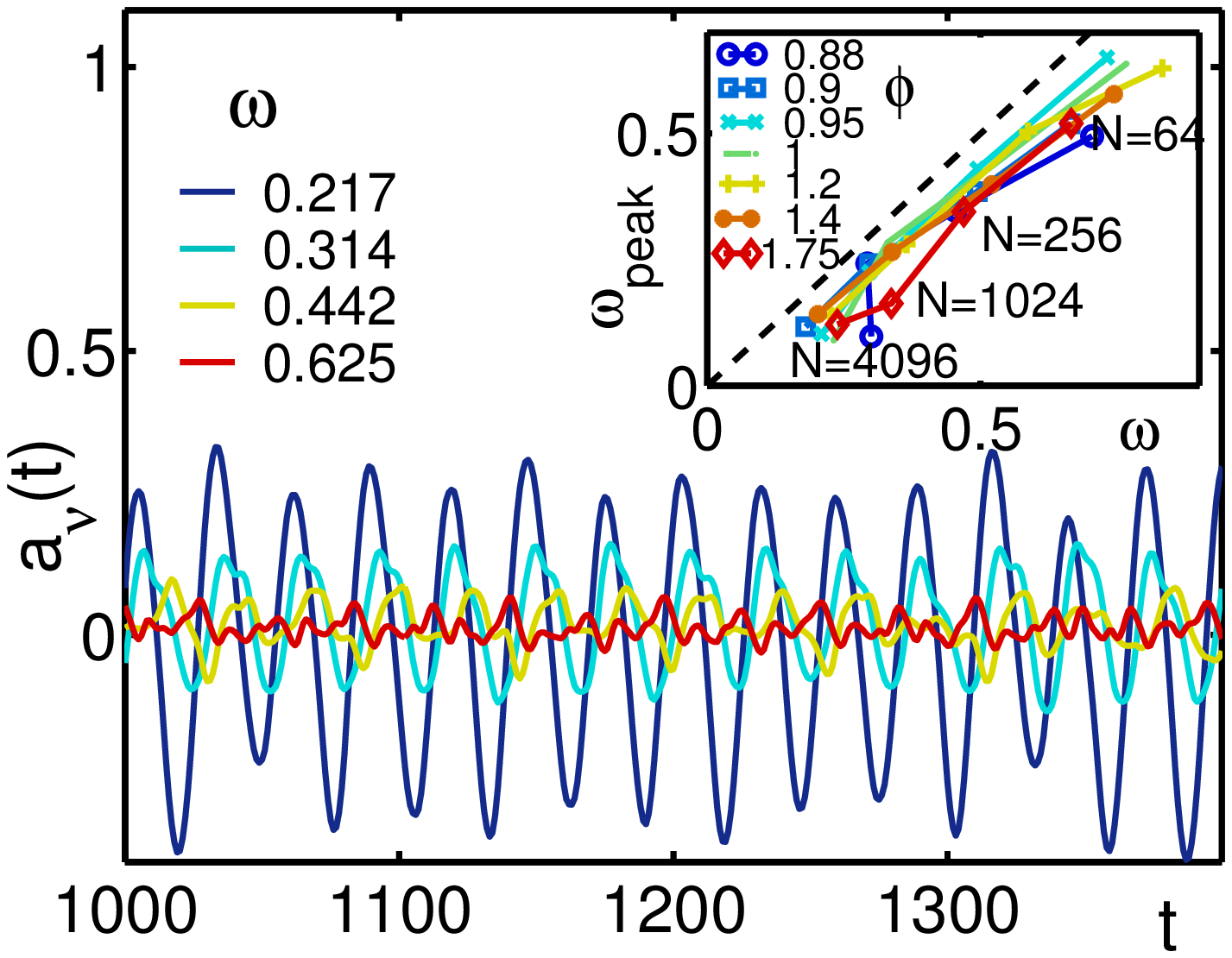}
\caption{(color online) Left: Density of states of the mean positions (dashed) and time-averaged projection of the displacements on the modes (solid, scaled for visibility) for $\phi=0.86$. Inset: Scaling of peak with system size $N_t$ at $v_{0}=0.025$; the line has a slope of $-0.3$. Right: Time projection of the displacement on the modes $a_{\nu}(t)$ for four representative modes in the undamped and damped region ($\phi=0.86$, $v_0=0.025$). Inset: Measured oscillation frequency $\omega'$ as a function of $\omega$.}
\label{fig:DOS_projection}
\end{figure}
Confinement plays a crucial role here.  For an open system or one with with periodic boundary conditions, the lowest frequency modes  are  the two translations allowed by symmetry, and $\omega_{\text{min}}=0$.  At high density the system is then in the aligned phase, with $v_{rms}=v_{0} \Delta$.
For a confined system, the lowest frequency modes are set by the system size.

We test these predictions in the jammed phase by numerically calculating the normal modes of the mean particle positions. The resulting density of states is indistinguishable from the known form for jammed packings at the same density (see Figure~\ref{fig:DOS_projection}, left), except for the absence of the two translational modes and a handfull of negative eigenvalues (5-10 for $N_t=1000$) which hint at rearrangements in the oscillatory phase. We have confirmed this by equilibrating the mean positions at $v_{0}=0$, which leads to a small number of rearrangements. This observation is likely linked to the extremely limited range of linear response in jammed packings.

For both sets of mean positions (at finite and vanishing $v_0$), we project the motion of the system on the modes. After a transient,  the $a_{\nu}(t)$ corresponding to the  lowest-frequency modes are undamped and oscillate at a single frequency $\omega'$. In contrast, the $a_\nu(t)$ of the higher frequency modes describe strongly damped forced oscillations at the same $\omega'$ (see Fig.~\ref{fig:DOS_projection}, right). In Fig.~\ref{fig:DOS_projection} (left), we show the time-averaged mean-square projection coefficient $P(\omega)$ as a function of frequency. It is dominated by a peak at the lowest frequencies where the DOS is finite. The peak frequency is independent of $v_{0}$, as predicted by Eq.~\ref{eq:omega_theory}. The inset to Fig.~\ref{fig:DOS_projection}(left) shows the scaling of the peak frequency with system size. Finally, the inset to Fig.~\ref{fig:DOS_projection}(right), displays a direct  measurement of $\omega_{\nu}'$ as a function of the mode frequency. This was obtained by Fourier transforming $a_{\nu}(t)$ for individual modes and averaging the result over the $P(\omega)$-distribution. We find that the single frequency $\omega' \leq \omega_{\nu}$, consistent with $\omega'$ being set by the lowest modes ($\mu=\tau=1$ for our runs), and with higher-frequency modes doing forced oscillations at $\omega'$. Nonlinear effects can  be analyzed perturbatively and will be discussed in a future publication. 

We have studied numerically a model of self-propelled particles with soft repulsive interactions at high density, motivated by recent experiments on migrating cell layers.  The analysis of the role of adhesive forces, viscous interparticle interactions and  of the distribution of stress and strain in unconfined systems is currently underway.

\acknowledgments
This work was supported by the National Science Foundation through awards DMR-0806511 and DMR-1004789.  We thank Olivier Dauchot and Ben Burdick for useful discussions.
The computations were carried out on SUGAR, a computing
cluster supported by NSF-PHY-1040231.

\bibliography{biblio}

\begin{thebibliography}{10}%
\makeatletter
\providecommand \@ifxundefined [1]{%
 \ifx #1\undefined \expandafter \@firstoftwo
 \else \expandafter \@secondoftwo
\fi
}%
\providecommand \@ifnum [1]{%
 \ifnum #1\expandafter \@firstoftwo
 \else \expandafter \@secondoftwo
\fi
}%
\providecommand \enquote [1]{``#1''}%
\providecommand \bibnamefont  [1]{#1}%
\providecommand \bibfnamefont [1]{#1}%
\providecommand \citenamefont [1]{#1}%
\providecommand\href[0]{\@sanitize\@href}%
\providecommand\@href[1]{\endgroup\@@startlink{#1}\endgroup\@@href}%
\providecommand\@@href[1]{#1\@@endlink}%
\providecommand \@sanitize [0]{\begingroup\catcode`\&12\catcode`\#12\relax}%
\@ifxundefined \pdfoutput {\@firstoftwo}{%
 \@ifnum{\z@=\pdfoutput}{\@firstoftwo}{\@secondoftwo}%
}{%
 \providecommand\@@startlink[1]{\leavevmode\special{html:<a href="#1">}}%
 \providecommand\@@endlink[0]{\special{html:</a>}}%
}{%
 \providecommand\@@startlink[1]{%
  \leavevmode
  \pdfstartlink
   attr{/Border[0 0 1 ]/H/I/C[0 1 1]}%
   user{/Subtype/Link/A<</Type/Action/S/URI/URI(#1)>>}%
  \relax
 }%
 \providecommand\@@endlink[0]{\pdfendlink}%
}%
\providecommand \url  [0]{\begingroup\@sanitize \@url }%
\providecommand \@url [1]{\endgroup\@href {#1}{\urlprefix}}%
\providecommand \urlprefix [0]{URL }%
\providecommand \Eprint[0]{\href }%
\@ifxundefined \urlstyle {%
  \providecommand \doi [1]{doi:\discretionary{}{}{}#1}%
}{%
  \providecommand \doi [0]{doi:\discretionary{}{}{}\begingroup
  \urlstyle{rm}\Url }%
}%
\providecommand \doibase [0]{http://dx.doi.org/}%
\providecommand \Doi[1]{\href{\doibase#1}}%
\providecommand \bibAnnote [3]{%
  \BibitemShut{#1}%
  \begin{quotation}\noindent
    \textsc{Key:}\ #2\\\textsc{Annotation:}\ #3%
  \end{quotation}%
}%
\providecommand \bibAnnoteFile [2]{%
  \IfFileExists{#2}{\bibAnnote {#1} {#2} {\input{#2}}}{}%
}%
\providecommand \typeout [0]{\immediate \write \m@ne }%
\providecommand \selectlanguage [0]{\@gobble}%
\providecommand \bibinfo [0]{\@secondoftwo}%
\providecommand \bibfield [0]{\@secondoftwo}%
\providecommand \translation [1]{[#1]}%
\providecommand \BibitemOpen[0]{}%
\providecommand \bibitemStop [0]{}%
\providecommand \bibitemNoStop [0]{.\EOS\space}%
\providecommand \EOS [0]{\spacefactor3000\relax}%
\providecommand \BibitemShut [1]{\csname bibitem#1\endcsname}%
\bibitem{Angelini2011}%
  \BibitemOpen
  \bibfield{author}{%
  \bibinfo {author} {\bibfnamefont{T.~E.}\ \bibnamefont{Angelini}}, \bibinfo
  {author} {\bibfnamefont{E.}~\bibnamefont{Hannezo}}, \bibinfo {author}
  {\bibfnamefont{X.}~\bibnamefont{Trepat}}, \bibinfo {author}
  {\bibfnamefont{M.}~\bibnamefont{Marquez}}, \bibinfo {author}
  {\bibfnamefont{J.~J.}\ \bibnamefont{Fredberg}},\ and\ \bibinfo {author}
  {\bibfnamefont{D.~A.}\ \bibnamefont{Weitz}},\ }%
  \bibfield{journal}{%
  \bibinfo {journal} {PNAS}\ }%
  \textbf{\bibinfo {volume} {108}},\ \bibinfo {pages} {4714} (\bibinfo {year}
  {2011})%
  \bibAnnoteFile{NoStop}{Angelini2011}%
\bibitem{Angelini2010}%
  \BibitemOpen
  \bibfield{author}{%
  \bibinfo {author} {\bibfnamefont{T.~E.}\ \bibnamefont{Angelini}}, \bibinfo
  {author} {\bibfnamefont{E.}~\bibnamefont{Hannezo}}, \bibinfo {author}
  {\bibfnamefont{X.}~\bibnamefont{Trepat}}, \bibinfo {author}
  {\bibfnamefont{J.~J.}\ \bibnamefont{Fredberg}},\ and\ \bibinfo {author}
  {\bibfnamefont{D.~A.}\ \bibnamefont{Weitz}},\ }%
  \bibfield{journal}{%
  \bibinfo {journal} {Phys. Rev. Lett.}\ }%
  \textbf{\bibinfo {volume} {104}},\ \bibinfo {pages} {168104} (\bibinfo {year}
  {2010})%
  \bibAnnoteFile{NoStop}{Angelini2010}%
\bibitem{Trepat2009}%
  \BibitemOpen
  \bibfield{author}{%
  \bibinfo {author} {\bibfnamefont{X.}~\bibnamefont{Trepat}}, \bibinfo {author}
  {\bibfnamefont{M.~R.}\ \bibnamefont{Wasserman}}, \bibinfo {author}
  {\bibfnamefont{T.~E.}\ \bibnamefont{Angelini}}, \bibinfo {author}
  {\bibfnamefont{E.}~\bibnamefont{Millet}}, \bibinfo {author}
  {\bibfnamefont{D.~A.}\ \bibnamefont{Weitz}}, \bibinfo {author}
  {\bibfnamefont{J.~P.}\ \bibnamefont{Butler}},\ and\ \bibinfo {author}
  {\bibfnamefont{J.~J.}\ \bibnamefont{Fredberg}},\ }%
  \bibfield{journal}{%
  \bibinfo {journal} {Nat. Phys.}\ }%
  \textbf{\bibinfo {volume} {5}},\ \bibinfo {pages} {426} (\bibinfo {year}
  {2009})%
  \bibAnnoteFile{NoStop}{Trepat2009}%
\bibitem{Poujade2007}%
  \BibitemOpen
  \bibfield{author}{%
  \bibinfo {author} {\bibfnamefont{M.}~\bibnamefont{Poujade}}, \bibinfo
  {author} {\bibfnamefont{E.}~\bibnamefont{Grasland-Mongrain}}, \bibinfo
  {author} {\bibfnamefont{A.}~\bibnamefont{Hertzog}}, \bibinfo {author}
  {\bibfnamefont{J.}~\bibnamefont{Jouanneau}}, \bibinfo {author}
  {\bibfnamefont{P.}~\bibnamefont{Chavrier}}, \bibinfo {author}
  {\bibfnamefont{B.}~\bibnamefont{Ladoux}}, \bibinfo {author}
  {\bibfnamefont{A.}~\bibnamefont{Buguin}},\ and\ \bibinfo {author}
  {\bibfnamefont{P.}~\bibnamefont{Silberzan}},\ }%
  \bibfield{journal}{%
  \bibinfo {journal} {PNAS}\ }%
  \textbf{\bibinfo {volume} {104}},\ \bibinfo {pages} {15988} (\bibinfo {year}
  {2007})%
  \bibAnnoteFile{NoStop}{Poujade2007}%
\bibitem{Petitjean2010}%
  \BibitemOpen
  \bibfield{author}{%
  \bibinfo {author} {\bibfnamefont{L.}~\bibnamefont{Petitjean}}, \bibinfo
  {author} {\bibfnamefont{M.}~\bibnamefont{Reffay}}, \bibinfo {author}
  {\bibfnamefont{E.}~\bibnamefont{Grasland-Mongrain}}, \bibinfo {author}
  {\bibfnamefont{M.}~\bibnamefont{Poujade}}, \bibinfo {author}
  {\bibfnamefont{B.}~\bibnamefont{Ladoux}}, \bibinfo {author}
  {\bibfnamefont{A.}~\bibnamefont{Buguin}},\ and\ \bibinfo {author}
  {\bibfnamefont{P.}~\bibnamefont{Silberzan}},\ }%
  \bibfield{journal}{%
  \bibinfo {journal} {Biophys. J.}\ }%
  \textbf{\bibinfo {volume} {98}},\ \bibinfo {pages} {1790} (\bibinfo {year}
  {2010})%
  \bibAnnoteFile{NoStop}{Petitjean2010}%
\bibitem{Narayan2007}%
  \BibitemOpen
  \bibfield{author}{%
  \bibinfo {author} {\bibfnamefont{V.}~\bibnamefont{Narayan}}, \bibinfo
  {author} {\bibfnamefont{S.}~\bibnamefont{Ramaswamy}},\ and\ \bibinfo {author}
  {\bibfnamefont{N.}~\bibnamefont{Menon}},\ }%
  \bibfield{journal}{%
  \bibinfo {journal} {Science}\ }%
  \textbf{\bibinfo {volume} {317}},\ \bibinfo {pages} {105} (\bibinfo {year}
  {2007})%
  \bibAnnoteFile{NoStop}{Narayan2007}%
\bibitem{Deseigne2010}%
  \BibitemOpen
  \bibfield{author}{%
  \bibinfo {author} {\bibfnamefont{J.}~\bibnamefont{Deseigne}}, \bibinfo
  {author} {\bibfnamefont{O.}~\bibnamefont{Dauchot}},\ and\ \bibinfo {author}
  {\bibfnamefont{H.}~\bibnamefont{Chat\'e}},\ }%
  \bibfield{journal}{%
  \bibinfo {journal} {Phys. Rev. Lett.}\ }%
  \textbf{\bibinfo {volume} {105}},\ \bibinfo {pages} {098001} (\bibinfo {year}
  {2010})%
  \bibAnnoteFile{NoStop}{Deseigne2010}%
\bibitem{Helbing2007}%
  \BibitemOpen
  \bibfield{author}{%
  \bibinfo {author} {\bibfnamefont{D.}~\bibnamefont{Helbing}}, \bibinfo
  {author} {\bibfnamefont{A.}~\bibnamefont{Johansson}},\ and\ \bibinfo {author}
  {\bibfnamefont{H.~Z.}\ \bibnamefont{Al-Abideen}},\ }%
  \bibfield{journal}{%
  \bibinfo {journal} {Phys. Rev. E}\ }%
  \textbf{\bibinfo {volume} {75}},\ \bibinfo {pages} {046109} (\bibinfo {year}
  {2007})%
  \bibAnnoteFile{NoStop}{Helbing2007}%
\bibitem{Vicsek1995}%
  \BibitemOpen
  \bibfield{author}{%
  \bibinfo {author} {\bibfnamefont{T.}~\bibnamefont{Vicsek}}, \bibinfo {author}
  {\bibfnamefont{A.}~\bibnamefont{Czir\'ok}}, \bibinfo {author}
  {\bibfnamefont{E.}~\bibnamefont{Ben-Jacob}}, \bibinfo {author}
  {\bibfnamefont{I.}~\bibnamefont{Cohen}},\ and\ \bibinfo {author}
  {\bibfnamefont{O.}~\bibnamefont{Shochet}},\ }%
  \bibfield{journal}{%
  \bibinfo {journal} {Phys. Rev. Lett.}\ }%
  \textbf{\bibinfo {volume} {75}},\ \bibinfo {pages} {1226} (\bibinfo {year}
  {1995})%
  \bibAnnoteFile{NoStop}{Vicsek1995}%
\bibitem{Gregoire2004}%
  \BibitemOpen
  \bibfield{author}{%
  \bibinfo {author} {\bibfnamefont{G.}~\bibnamefont{Gr\'egoire}}\ and\ \bibinfo
  {author} {\bibfnamefont{H.}~\bibnamefont{Chat\'e}},\ }%
  \bibfield{journal}{%
  \bibinfo {journal} {Phys. Rev. Lett.}\ }%
  \textbf{\bibinfo {volume} {92}},\ \bibinfo {pages} {025702} (\bibinfo {year}
  {2004})%
  \bibAnnoteFile{NoStop}{Gregoire2004}%
\bibitem{Chate2008}%
  \BibitemOpen
  \bibfield{author}{%
  \bibinfo {author} {\bibfnamefont{H.}~\bibnamefont{Chat\'e}}, \bibinfo
  {author} {\bibfnamefont{F.}~\bibnamefont{Ginelli}}, \bibinfo {author}
  {\bibfnamefont{G.}~\bibnamefont{Gr\'egoire}}, \bibinfo {author}
  {\bibfnamefont{F.}~\bibnamefont{Peruani}},\ and\ \bibinfo {author}
  {\bibfnamefont{F.}~\bibnamefont{Raynaud}},\ }%
  \bibfield{journal}{%
  \bibinfo {journal} {Eur. Phys. J. B}\ }%
  \textbf{\bibinfo {volume} {64}},\ \bibinfo {pages} {451} (\bibinfo {year}
  {2008})%
  \bibAnnoteFile{NoStop}{Chate2008}%
\bibitem{Chate2006}%
  \BibitemOpen
  \bibfield{author}{%
  \bibinfo {author} {\bibfnamefont{H.}~\bibnamefont{Chat\'e}}, \bibinfo
  {author} {\bibfnamefont{F.}~\bibnamefont{Ginelli}},\ and\ \bibinfo {author}
  {\bibfnamefont{R.}~\bibnamefont{Montagne}},\ }%
  \bibfield{journal}{%
  \bibinfo {journal} {Phys. Rev. Lett.}\ }%
  \textbf{\bibinfo {volume} {96}},\ \bibinfo {pages} {180602} (\bibinfo {year}
  {2006})%
  \bibAnnoteFile{NoStop}{Chate2006}%
\bibitem{TonerTuRamaswamy2005}%
  \BibitemOpen
  \bibfield{author}{%
  \bibinfo {author} {\bibfnamefont{J.}~\bibnamefont{Toner}}, \bibinfo {author}
  {\bibfnamefont{Y.}~\bibnamefont{Tu}},\ and\ \bibinfo {author}
  {\bibfnamefont{S.}~\bibnamefont{Ramaswamy}},\ }%
  \bibfield{journal}{%
  \bibinfo {journal} {Ann. Phys.}\ }%
  \textbf{\bibinfo {volume} {318}},\ \bibinfo {pages} {170} (\bibinfo {year}
  {2005})%
  \bibAnnoteFile{NoStop}{TonerTuRamaswamy2005}%
\bibitem{Liu1998}%
  \BibitemOpen
  \bibfield{author}{%
  \bibinfo {author} {\bibfnamefont{A.~J.}\ \bibnamefont{Liu}}\ and\ \bibinfo
  {author} {\bibfnamefont{S.~R.}\ \bibnamefont{Nagel}},\ }%
  \bibfield{journal}{%
  \bibinfo {journal} {Nature}\ }%
  \textbf{\bibinfo {volume} {396}},\ \bibinfo {pages} {21} (\bibinfo {year}
  {1998})%
  \bibAnnoteFile{NoStop}{Liu1998}%
\bibitem{Berthier2009}%
  \BibitemOpen
  \bibfield{author}{%
  \bibinfo {author} {\bibfnamefont{L.}~\bibnamefont{Berthier}}\ and\ \bibinfo
  {author} {\bibfnamefont{T.~A.}\ \bibnamefont{Witten}},\ }%
  \bibfield{journal}{%
  \bibinfo {journal} {Phys. Rev. E}\ }%
  \textbf{\bibinfo {volume} {80}},\ \bibinfo {pages} {021502} (\bibinfo {year}
  {2009})%
  \bibAnnoteFile{NoStop}{Berthier2009}%
\bibitem{Weeks2000}%
  \BibitemOpen
  \bibfield{author}{%
  \bibinfo {author} {\bibfnamefont{E.~R.}\ \bibnamefont{Weeks}}, \bibinfo
  {author} {\bibfnamefont{J.~C.}\ \bibnamefont{Crocker}}, \bibinfo {author}
  {\bibfnamefont{A.~C.}\ \bibnamefont{Lewitt}}, \bibinfo {author}
  {\bibfnamefont{A.}~\bibnamefont{Schofield}},\ and\ \bibinfo {author}
  {\bibfnamefont{D.~A.}\ \bibnamefont{Weitz}},\ }%
  \bibfield{journal}{%
  \bibinfo {journal} {Science}\ }%
  \textbf{\bibinfo {volume} {287}},\ \bibinfo {pages} {621} (\bibinfo {year}
  {2000})%
  \bibAnnoteFile{NoStop}{Weeks2000}%
\bibitem{O'Hern2003}%
  \BibitemOpen
  \bibfield{author}{%
  \bibinfo {author} {\bibfnamefont{C.~S.}\ \bibnamefont{O'Hern}}, \bibinfo
  {author} {\bibfnamefont{L.~E.}\ \bibnamefont{Silbert}}, \bibinfo {author}
  {\bibfnamefont{A.~J.}\ \bibnamefont{Liu}},\ and\ \bibinfo {author}
  {\bibfnamefont{S.~R.}\ \bibnamefont{Nagel}},\ }%
  \bibfield{journal}{%
  \bibinfo {journal} {Phys. Rev. E}\ }%
  \textbf{\bibinfo {volume} {68}},\ \bibinfo {pages} {011306} (\bibinfo {year}
  {2003})%
  \bibAnnoteFile{NoStop}{O'Hern2003}%
\bibitem{Silbert2005}%
  \BibitemOpen
  \bibfield{author}{%
  \bibinfo {author} {\bibfnamefont{L.~E.}\ \bibnamefont{Silbert}}, \bibinfo
  {author} {\bibfnamefont{A.~J.}\ \bibnamefont{Liu}},\ and\ \bibinfo {author}
  {\bibfnamefont{S.~R.}\ \bibnamefont{Nagel}},\ }%
  \bibfield{journal}{%
  \bibinfo {journal} {Phys. Rev. Lett.}\ }%
  \textbf{\bibinfo {volume} {95}},\ \bibinfo {pages} {098301} (\bibinfo {year}
  {2005})%
  \bibAnnoteFile{NoStop}{Silbert2005}%
\bibitem{Maloney2006}%
  \BibitemOpen
  \bibfield{author}{%
  \bibinfo {author} {\bibfnamefont{C.~E.}\ \bibnamefont{Maloney}},\ }%
  \bibfield{journal}{%
  \bibinfo {journal} {Phys. Rev. Lett.}\ }%
  \textbf{\bibinfo {volume} {97}},\ \bibinfo {pages} {035503} (\bibinfo {year}
  {2006})%
  \bibAnnoteFile{NoStop}{Maloney2006}%
\bibitem{Brito2007}%
  \BibitemOpen
  \bibfield{author}{%
  \bibinfo {author} {\bibfnamefont{C.}~\bibnamefont{Brito}}\ and\ \bibinfo
  {author} {\bibfnamefont{M.}~\bibnamefont{Wyart}},\ }%
  \bibfield{journal}{%
  \bibinfo {journal} {J. Stat. Mech}\ }%
  \textbf{\bibinfo {volume} {L08003}} (\bibinfo {year} {2007})%
  \bibAnnoteFile{NoStop}{Brito2007}%
\bibitem{Szabo2006}%
  \BibitemOpen
  \bibfield{author}{%
  \bibinfo {author} {\bibfnamefont{B.}~\bibnamefont{Szab\'o}}, \bibinfo
  {author} {\bibfnamefont{G.~J.}\ \bibnamefont{Szollosi}}, \bibinfo {author}
  {\bibfnamefont{B.}~\bibnamefont{Gonci}}, \bibinfo {author}
  {\bibfnamefont{Z.}~\bibnamefont{Juranyi}}, \bibinfo {author}
  {\bibfnamefont{D.}~\bibnamefont{Selmeczi}},\ and\ \bibinfo {author}
  {\bibfnamefont{T.}~\bibnamefont{Vicsek}},\ }%
  \bibfield{journal}{%
  \bibinfo {journal} {Phys. Rev. E}\ }%
  \textbf{\bibinfo {volume} {74}},\ \bibinfo {pages} {061908} (\bibinfo {year}
  {2006})%
  \bibAnnoteFile{NoStop}{Szabo2006}%
\bibitem{Bindschadler2007}%
  \BibitemOpen
  \bibfield{author}{%
  \bibinfo {author} {\bibfnamefont{M.}~\bibnamefont{Bindschadler}}\ and\
  \bibinfo {author} {\bibfnamefont{J.~L.}\ \bibnamefont{Grath}},\ }%
  \bibfield{journal}{%
  \bibinfo {journal} {J. Cell Sci.}\ }%
  \textbf{\bibinfo {volume} {120}},\ \bibinfo {pages} {876} (\bibinfo {year}
  {2007})%
  \bibAnnoteFile{NoStop}{Bindschadler2007}%
\bibitem{Szabo2010}%
  \BibitemOpen
  \bibfield{author}{%
  \bibinfo {author} {\bibfnamefont{A.}~\bibnamefont{Szab\'o}}, \bibinfo
  {author} {\bibfnamefont{R.}~\bibnamefont{\'Unnep}}, \bibinfo {author}
  {\bibfnamefont{E.}~\bibnamefont{M\'ehes}}, \bibinfo {author}
  {\bibfnamefont{W.~O.}\ \bibnamefont{Twal}}, \bibinfo {author}
  {\bibfnamefont{W.~S.}\ \bibnamefont{Argraves}}, \bibinfo {author}
  {\bibfnamefont{Y.}~\bibnamefont{Cao}},\ and\ \bibinfo {author}
  {\bibfnamefont{A.}~\bibnamefont{Czir\'ok}},\ }%
  \bibfield{journal}{%
  \bibinfo {journal} {Phys. Biol.}\ }%
  \textbf{\bibinfo {volume} {7}},\ \bibinfo {pages} {046007} (\bibinfo {year}
  {2010})%
  \bibAnnoteFile{NoStop}{Szabo2010}%
\bibitem{Howse2010}%
  \BibitemOpen
  \bibfield{author}{%
  \bibinfo {author} {\bibfnamefont{J.~R.}\ \bibnamefont{Howse}}, \bibinfo
  {author} {\bibfnamefont{R.~A.~L.}\ \bibnamefont{Jones}}, \bibinfo {author}
  {\bibfnamefont{A.~J.}\ \bibnamefont{Ryan}}, \bibinfo {author}
  {\bibfnamefont{T.}~\bibnamefont{Gough}}, \bibinfo {author}
  {\bibfnamefont{R.}~\bibnamefont{Vafabakhsh}},\ and\ \bibinfo {author}
  {\bibfnamefont{R.}~\bibnamefont{Golestanian}},\ }%
  \bibfield{journal}{%
  \bibinfo {journal} {Phys. Rev. Lett.}\ }%
  \textbf{\bibinfo {volume} {99}},\ \bibinfo {pages} {048102} (\bibinfo {year}
  {2007})%
  \bibAnnoteFile{NoStop}{Howse2010}%
\bibitem{Zhang2010}%
  \BibitemOpen
  \bibfield{author}{%
  \bibinfo {author} {\bibfnamefont{H.~P.}\ \bibnamefont{Zhang}}, \bibinfo
  {author} {\bibfnamefont{A.}~\bibnamefont{Be'er}}, \bibinfo {author}
  {\bibfnamefont{E.-L.}\ \bibnamefont{Florin}},\ and\ \bibinfo {author}
  {\bibfnamefont{H.~L.}\ \bibnamefont{Swinney}},\ }%
  \bibfield{journal}{%
  \bibinfo {journal} {PNAS}\ }%
  \textbf{\bibinfo {volume} {107}},\ \bibinfo {pages} {13626} (\bibinfo {year}
  {2010})%
  \bibAnnoteFile{NoStop}{Zhang2010}%
\end{thebibliography}%

\end{document}